%% file: main.tex
\documentclass{article}
\usepackage{spconf,amsmath,graphicx,url}
\usepackage{CJKutf8}
\usepackage{hyperref}

\input{local_definition}

\title{Coco-Nut: Corpus of Japanese Utterance and Voice Characteristics Description for Prompt-based Control}
\name{Aya Watanabe, Shinnosuke Takamichi, Yuki Saito, Wataru Nakata, Detai Xin, Hiroshi Saruwatari\thanks{This work is supported by JSPS KAKENHI 21H04900, 22H03639, 23H03418 (practical experiment), JST FOREST JPMJFR226V, and Moonshot R\&D Grant Number JPMJPS2011 (algorithm development). We also appreciate Takaaki Saeki and Yuta Matsunaga of the University of Tokyo for their help.}}
\address{The University of Tokyo, Japan.}
\begin{document}
\ninept
\maketitle
\input{section/abstract}
\input{section/introduction}
\input{section/related-work}
\input{section/corpus-construction}
%
\input{section/experiment}
\input{section/conclusion}
\newpage
\bibliographystyle{bib/IEEEbib}
\bibliography{bib/refs}

\end{document}

%% file: local_definition.tex
\newcommand{\Fig}[1]{Figure~\ref{fig:#1}} 
\newcommand{\Sec}[1]{Section~\ref{sec:#1}} 

\newcommand{\jp}[1]{\begin{CJK}{UTF8}{ipxm}#1\end{CJK}} 

\newcommand{\drawfig}[4]{ 
    \begin{figure}[#1]
    \centering 
    \vspace{0mm}
    \includegraphics[width=#2]{#3} 
    \vspace{-4mm}
    \caption{#4}
    \label{fig:#3}
    \vspace{-4mm}
    \end{figure}
}

%% file: section/abstract.tex
\begin{abstract} \vspace{-1mm}
  In text-to-speech, controlling voice characteristics is important in achieving various-purpose speech synthesis. Considering the success of text-conditioned generation, such as text-to-image, free-form text instruction should be useful for intuitive and complicated control of voice characteristics. A sufficiently large corpus of high-quality and diverse voice samples with corresponding free-form descriptions can advance such control research. However, neither an open corpus nor a scalable method is currently available.
  To this end, we develop Coco-Nut, a new corpus including diverse Japanese utterances, along with text transcriptions and free-form voice characteristics descriptions.
  Our methodology to construct this corpus consists of 1) automatic collection of voice-related audio data from the Internet, 2) quality assurance, and 3) manual annotation using crowdsourcing. Additionally, we benchmark our corpus on the prompt embedding model trained by contrastive speech-text learning.
\end{abstract}

%% file: section/introduction.tex
\vspace{-3mm}
\section{Introduction}
\vspace{-2mm}
\label{sec:intro}

    In human speech production, the speaker's voice carries not only linguistic content but also unique vocal characteristics. Text-to-speech (TTS) tasks that imitate the speech production involve two significant challenges: synthesizing highly intelligible speech from the provided text (referred to as ``content prompt'' in this paper) and controlling the voice characteristics. This is because the characteristics greatly influence the listener's perception, affecting their understanding of the speaker's personality, emotion, and overall impression. Several methods of voice characteristics control have been proposed, 
    such as a speaker index~\cite{hojo16speakercode}, 
    speaker attributes~\cite{stanton2022speaker,watanabe22mid-attribute-speaker-generation}, 
    personality~\cite{gustafson21personality-in-the-mix}, 
    and so on~\cite{zhang19learning-to-speak,rui21reinforcement-learning-emotional-tts,ohta10voice-quality-gmm-vc,raitio20controllable-tts}.
    However, these methods only enable control over a narrow and simplistic range of voice characteristics, limiting their applicability in various contexts.

    There has been significant advancement in techniques for synthesizing media using free-form text descriptions (text prompts). This progress is evident in various fields, such as text-to-image~\cite{ramesh21dalle}, text-to-audio~\cite{elizalde22clap}, text-to-music~\cite{huang22mulan}, and text-to-video~\cite{ho22imagenvideo}. The potential of prompt-based media generation is to manipulate complicated media components, with benefits accruing from the ongoing advancements in large language models (LLMs)~\cite{radford21clip,elizalde22clap}. Following these trends, we believe that voice characteristics control by a free-form description opens new doors for TTS tasks. Hence, our goal is to develop TTS capable of controlling vocal characteristics through free-form descriptions, leading to the construction of a dedicated corpus. We refer to this free-form description and TTS synthesizer as the ``characteristics prompt'' and ``Prompt TTS,'' respectively. As depicted in \Fig{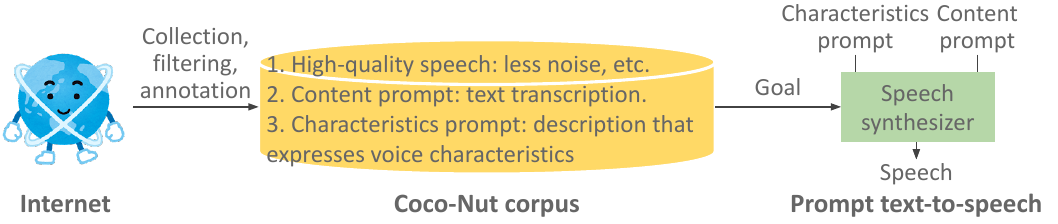}, the aim of Prompt TTS is to synthesize speech that aligns with the prompted linguistic content and voice characteristics. The corpus designed for this purpose should encompass a wide array of vocal characteristics, unlike the existing TTS corpora~\cite{zen19libritts,takamichi2020jsut} which tend to cover only a limited range of voice attributes. However, neither an open corpus nor a scalable methodology to construct the corpus is currently available.

    \drawfig{t}{\linewidth}{figure/overview.pdf}
    {Our Coco-Nut corpus towards prompt TTS. Characteristics prompt and content prompt are, for example, ``middle-aged man's voice speaking in a clear and polite tone'' and ``Welcome to our office!'' respectively. Speech synthesizer synthesizes speech of the prompted content with the prompted voice characteristics.}
    
    In this paper, we propose a methodology for constructing a corpus toward Prompt TTS. Our methodology consists of 1) machine-learning-based automatic collection of voice-related audio data from the Internet, 2) quality assurance to enhance the quality of content prompts and speech in the corpus, and 3) manual annotation of characteristics prompt using crowdsourcing. With this methodology, we construct an open corpus, \textit{Coco-Nut}\footnote{\textbf{Co}rpus of \textbf{co}nnecting \textbf{N}ihongo \textbf{u}tterance and \textbf{t}ext. ``Nihongo'' means the Japanese language in Japanese.}, which is available at our project page\footnote{\url{https://sites.google.com/site/shinnosuketakamichi/research-topics/coconut_corpus}}. 
    This paper also benchmarks our Coco-Nut corpus. The Coco-Nut corpus is used for training a contrastive speech-text training model that embeds characteristics prompt and speech into a same space. Experimental evaluation gives results of the corpus construction and performance of the benchmark system both on objective and subjective aspects.

%% file: section/related-work.tex
\vspace{-3mm}
\section{Related work} \vspace{-2mm} \label{sec:related-work}
    \subsection{Dataset for text-to-image}  \vspace{-2mm}\label{sec:related-work_text-image}
        Model training for text-to-image requires pairs of an image and text prompt that describes the image content.
        DALL-E~\cite{ramesh21dalle}, known as a pioneer in text-to-image, is trained using the MS-COCO dataset~\cite{lin14ms-coco} (image captioning dataset) and web data~\cite{sharma18conceptualcaptions}. 
        MS-COCO is a dataset used for image captioning, which involves manual annotation of texts that describe the image content. In addition to MS-COCO, the use of diverse data from the Internet in training significantly contributes to the synthesis of diverse images~\cite{ramesh21dalle}. Although HTML images and their accompanying alt-tag texts provide a massive amount of text-image pairs, data filtering is necessary due to noisiness of the Internet data. The pre-trained CLIP (contrastive language-image pretraining) model~\cite{radford21clip} is often used for data filtering purposes. The importance of data diversity and contrastive learning should be considered in other generation tasks, e.g., voice characteristics in this paper.

    \vspace{-2mm}
    \subsection{Dataset for text-to-audio and text-to-music}  \vspace{-2mm}\label{sec:related-work_text-audio}
        As the same to text-to-image, datasets for captioning are also available for text-to-audio.
        The typical examplpes are AudioCaps~\cite{kim19audiocaps} and Clotho~\cite{drossos20clotho}. Additionally, the text-audio version of CLIP, CLAP (contrastive language-audio pretraining) model~\cite{elizalde22clap}, is also used for data filtering~\cite{wu22laion-audio} before the training. MuLan~\cite{huang22mulan} in text-to-music proposes a method of retrieving music videos on the web and builds a machine learning model to identify whether the text attached to the video describes the music. This methodology has the potential to be applied to other than music.

        Unlike the text-to-audio and text-to-music cases, datasets for Prompt TTS are very limited\footnote{Audio captioning datasets~\cite{kim19audiocaps,drossos20clotho} include human voices as an environmental sound, but the voices do not strongly specify linguistic content.}. Existing studies have added characteristics prompts to small in-house and private datasets~\cite{guo22prompttts,guo23instructtts}. However, typical TTS corpora~\cite{zen19libritts,takamichi2020jsut} contain only limited voice characteristics. Considering the contribution of Internet data described in \Sec{related-work_text-image}, it is necessary to establish a methodology of corpus construction from the Internet data. Also, there is no open corpus that everyone can access.

    \vspace{-2mm}
    \subsection{Sequence generation from text}  \vspace{-2mm} \label{sec:related-work_text-representation}
        In sequence generation tasks such as text-to-video and text-to-audio, it is necessary to determine 1) \textit{overall concepts} that represent characteristics of the entire sequence and 2) \textit{sequence concepts} that represent characteristics of changes in the sequence. There are two kinds for describing these concepts using text. 
        
        The first is to describe both concepts in a single text, e.g., ``wooden figurine surfing on a surfboard in space''~\cite{ho22imagenvideo} in text-to-video and ``hip-hop features rap with an electronic backing''~\cite{huang22mulan} in text-to-music\footnote{MusicLM~\cite{agostinelli23musiclm} uses a variation of this kind by switching the description at fixed intervals (15 seconds in the paper) to allow for more fine-grained control of changes. This method is suitable for applications that generate sequences from rough descriptions.}. The second is to describe each concept in separate texts. Examples of this include ``bat hitting'' (overall concept) and ``ki-i-i-n'' (sequence concept) in text-to-audio~\cite{ohnaka22visual-onoma-to-wave}\footnote{LAION-Audio-630K~\cite{laion-ai-audio} uses text of overall concept for non-speech environmental sounds and that of sequence concept for speech-related environmental sounds.} and ``A toy fireman is lifting weights'' (sequence concept) in text-to-video~\cite{molad23dreamix}\footnote{Overall concept is given by an image in the paper.}. This kind of methods is suitable for applications that require fine-grained control over the sequence, such as TTS, where the linguistic content and voice characteristics are often controlled separately~\cite{guo22prompttts,guo23instructtts}. Therefore, we aim to collect content and characteristics prompts separately.

%% file: section/corpus-construction.tex
\vspace{-3mm}
\section{Corpus construction}  \vspace{-2mm} \label{sec:corpus-construction}
    \drawfig{t}{0.99\linewidth}{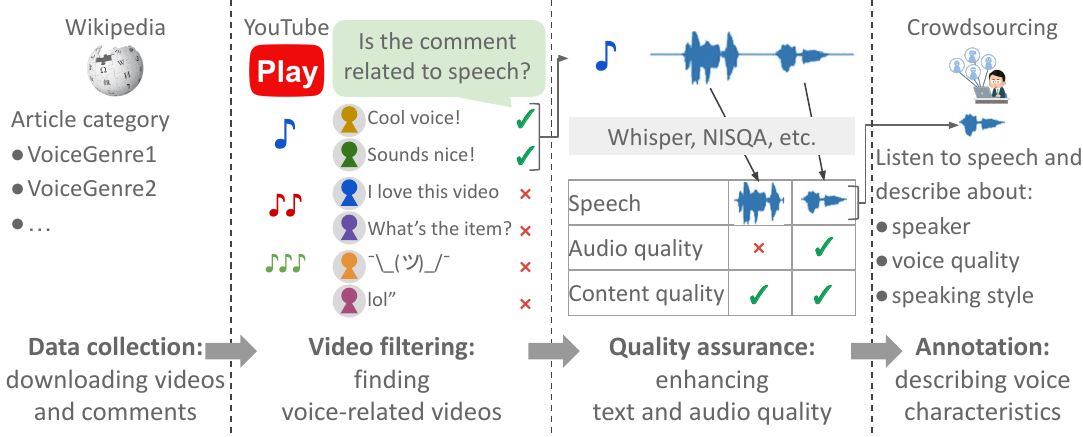}
    {Procedure of corpus construction.}
    
    \subsection{Corpus composition} \vspace{-2mm}
        The corpus for Prompt TTS should include:
        \begin{enumerate} \leftskip -4mm \itemsep -1mm \vspace{-1mm}
            \item \textbf{High-quality speech.} Speech data for TTS. Unlike data in speech-to-text corpora~\cite{chen2021gigaspeech,takamichi21jtubespeech,yin23reazonspeech}, it should be high-quality, e.g., less noise. Also, it is paired with the content prompt and characteristics prompt.
            \item \textbf{Content prompt.} Text transcriptions of speech. This corresponds to the sequence concept described in \Sec{related-work_text-representation}.
            \item \textbf{Characteristics prompt.} Free-form descriptions that express characteristics of speech. This corresponds to the overall concept described in \Sec{related-work_text-representation}.
        \end{enumerate} \vspace{-1mm}
        Existing approaches~\cite{guo22prompttts,guo23instructtts} for constructing this kind of corpora are to add characteristics prompts to existing TTS corpora consisting of high-quality speech and content prompts, e.g., \cite{takamichi2020jsut,zen19libritts}. However, as described in \Sec{related-work}, such corpora often lack diversity of voice characteristics. Therefore, we propose a methodology that builds a corpus from very noisy Internet data.

        Our methodology consists of the following four steps. \Fig{figure/procedure.pdf} illustrates these steps. Although the target language of this paper is Japanese, the process of these steps is language-independent; our methodology can be implemented in other languages than Japanese.
        \begin{enumerate} \leftskip -4mm \itemsep -1mm \vspace{-1mm}
            \item \textbf{Data collection.} Speech data candidates are searched out and obtained from the Internet. 
            \item \textbf{Video filtering.} Impressive voice data are filtered from the candidates. The ``impressive voice'' refers to those that have received a large number of responses on the Internet. Such data are expected to be characteristic voices and therefore suitable for the construction of the corpus.
            \item \textbf{Quality assurance.} Speech and its transcriptions (content prompts) are further filtered to guarantee quality of the corpus. 
            \item \textbf{Manual annotation.} Characteristics prompts are manually annotated to the speech data.
        \end{enumerate} \vspace{-1mm}
        The subsequent subsections describe the details of these steps.

    \vspace{-2mm}
    \subsection{Data collection} \vspace{-2mm}\label{sec:proposed-method-data-collection}
        To obtain speech data candidates, we make search phrases and input it into the search engine of video-sharing websites, e.g., YouTube. We select article categories related to speech from Wikipedia\footnote{For example, \url{https://en.wikipedia.org/wiki/List_of_YouTubers} in English.} in the target language and use the titles of Wikipedia articles belonging to those categories as search phrases. In addition, we add related phrases that are thought to be related to the search phrase (e.g., ``[article title] short clip''). After searching, we obtain a video ID, audio data, video title, and viewers' comments of the found videos.

    \vspace{-2mm}
    \subsection{Video filtering} \vspace{-2mm}
        By filtering through the above obtained video data, videos containing ``impressive voice'' are acquired. In this paper, we extract videos in which many people commented about voices in the videos. Two-stage filtering are conducted, and the voices of the filtered videos are forwarded to the next ``quality assurance'' step.

        \begin{enumerate} \leftskip -4mm \itemsep -1mm \vspace{-1mm}
            \item \textbf{Keyword matching-based pre-filtering.} 
            The obtained data contains many videos without audio or with nondescript voices. First, a rule-based video filter is applied. We use a set of keywords related to voice characteristics (e.g., ``listen'') to distinguish whether a viewer's comment on a video contain those keywords. If the number of comments containing the keywords in that video is greater than the threshold, the video is adopted.

            \item \textbf{Machine learning-based filtering.} 
            Machine learning is used to determine whether viewers' comments mention to the voice in the video, obtaining in videos with ``impressive voices''. We create training data for this machine learning. We randomly extract viewers' comments from videos and perform crowdsourcing-based annotations on the comment mentions. A title and comment of the video are presented to the crowdworkers\footnote{For example, ``Video title: My daily voice training method. Comment: Cool Voice!'' The answer will be ``1) related to speaking voice.'' Presenting title makes it easier for the crowdworkers to judge the comment content by having the crowdworkers imagine the content of the video.}. The crowdworkers answer whether the comment is 1) related to speaking voice, 2) related to singing voice, or 3) others. Before the annotation, we instruct crowdworkers that ``1)'' includes comments mentioning about the voice characteristics but does not include comments about the linguistic contents.

            A comment content classifier is trained using the above annotated data. The classifier model is BERT~\cite{devlin2018bert} followed by a linear layer. The input is a video title and comment, joined by ``[SEP]'' token that represents ``sentence separation'' in BERT. The output target is binary: 1) speech-related comment and 2) singing-related comment and others. 
            To improve classification performance, we decided to use the aforementioned keywords auxiliary. A subset of the keyword set was used, and only comments that matched one of the subsets were used to train and evaluate the classifier.

        \end{enumerate} \vspace{-1mm}



    \vspace{-2mm}
    \subsection{Quality assurance} \vspace{-2mm}
        Due to the collection of Internet data, there are some text and speech data samples that are of low quality and difficult to use. In order to ensure the quality of the audio data included in the corpus, the following processes are used to filter the data.

        \vspace{-2mm}
        \subsubsection{Audio quality} \vspace{-2mm}
        To ensure the quality of the sound, the following operations are performed.
        \begin{enumerate} \leftskip -4mm \itemsep -1mm \vspace{-1mm}
            \item 
            \textbf{Voice activity detection (VAD).}
            VAD is performed to extract only the segments containing voices from the entire video. We use inaSpeechSegmenter~\cite{ddoukhanicassp2018} to detect individual speech segments in the video.
            
            \item 
            \textbf{Denoising.} To enhance audio quality, we use Demucs\footnote{\url{https://github.com/facebookresearch/demucs}}, which is a powerful source separation model based on deep neural networks, to extract the voices from noise-contaminated voices.

            \item 
            \textbf{Audio quality assessment.} There is a variety of audio quality of speech, e.g., recording device quality and effective frequency band. Also, denoising process well eliminates background noise but sometimes drops speech component. To quantify quality degradation caused by these factors, we use NISQA~\cite{mittag21nisqa}, a multidimensional speech quality predictor. The NISQA score is calculated for each speech segment, and we filtered out the segments with the score lower than the pre-determined threshold\footnote{We found that speech component drop can be quantified by the NISQA score.}.
            
            \item 
            \textbf{Threshold for duration and audio volume.}
            We set the acceptable duration ranges to eliminate too long and too short voices. We also set the volume threshold and filtered out inaudible (low-volume) speech.
            

            \item 
            \textbf{Detection of multi-speaker voice and singing voice.}
            Data not intended for TTS, specifically singing voices and multi-speaker voices (e.g., cheering), are manually excluded.


            \item 
            \textbf{Voice characteristics variation.} 
            It is desirable for the corpus to include a variety of voice characteristics. To achieve this, we perform hierarchical clustering based on Ward's method~\cite{ward63clustering} using distances of $x$-vectors~\cite{snyder2018x}, which reflects not only voice quality but also speech style as suggested by~\cite{brown2021playing}. The $x$-vector is extracted for each speech segment by a pretrained $x$-vector extractor. Since speech segments with similar voice characteristics are expected to be grouped, we randomly sample one speech segment as the representative of each cluster. 
        \end{enumerate} \vspace{-1mm}

        \vspace{-2mm}
        \subsubsection{Content quality}  \vspace{-2mm}
        To select appropriate speech contents, the following processing steps are performed.
        
        \begin{enumerate} \leftskip -4mm \itemsep -1mm \vspace{-1mm}
            \item 
            \textbf{Speech-to-text and language identification.}
            To obtain content prompts of speech, we use pre-trained Whisper speech-to-text models~\cite{radford2022robust}. Jointly with speech-to-text, we identify language of speech by Whisper and filtered out speech of non-target language. Furthermore, manual identification is conducted to enhance the corpus quality\footnote{We found that language identification by Whisper alone would result in the inclusion of many voices of non-target languages.}.
            
            \item 
            \textbf{NSFW (not safe for work) word detection.} 
            We filtered out content prompts that include NSFW words. 
            We adopt keyword matching-based NSFW word detection; the text is filtered out if the lemmatized word is found in the NSFW word dictionary. Additional manual detection is conducted to enhance the corpus quality.
            
            \item 
            \textbf{Non-verbal voice detection.}
            Since TTS does not handle non-verbal voices, e.g., scream, we filter out non-verbal voices using a large language model and content prompt texts. Masked language model (MLM) scores~\cite{salazar20maskedlaugagemodel} based on BERT~\cite{devlin2018bert} are calculated for each segment's transcription. Since the masked tokens of content prompt text is highly predictable from the adjacent tokens\footnote{For example, let consider ``aa[MASK]aaaa,'' a partially masked content prompt of scream. The masked token ``[MASK]'' will be ``aa.''  }, the MLM score of the non-verbal voice becomes higher. We manually set a threshold against to the MLM score and filtered out speech with the MLM score higher than the threshold.
        \end{enumerate} \vspace{-1mm}

    \vspace{-2mm}
    \subsection{Manual annotation} \vspace{-2mm}
        Finally, we use crowdsourcing to add characteristics prompts to the collected voices. The employed crowdworkers listen to the presented voice and describe the voice characteristics. They are instructed to include speaker attributes, voice quality, and speaking style in their descriptions\footnote{The actual English-translated instruction is ``Describe what kinds of speaker (age, gender, etc.), voice quality (brisk, low voice, etc.), and speaking style (angry, fast, etc.) in a free-form description of at least 20 characters. Do not include the linguistic content of the speech, and do not use expressions that indicate personal likes and dislikes (e.g., my favorite voice and disliked way of speaking).''. }. Only descriptions with more than the threshold number of characters are accepted. 

        After collecting characteristics prompts, we manually filtered out characteristics prompts that include proper nouns and persons' name, e.g., ``The voice is similar to [celebrity's real name].'' This is done to prevent models trained on this corpus from generating the voices of actual individuals' name. We also perform text normalization to cleanse the descriptions.

        %
    
            %

%% file: section/experiment.tex
\vspace{-2mm}
\section{Experiments} \vspace{-2mm}\label{sec:experiment}
    \subsection{Data collection} \vspace{-2mm}
        The target language was Japanese. The data collection period was from July 2022 to March 2023. The number of comments per video was limited to the top 100 comments with the highest number of ``Likes.'' After extracting comments in the target language by rule-based language identification, comments with less than 3 characters or more than 50 characters were excluded. Table~\ref{tab:data-collection} lists the results of the data collection.

\input{table/data-collection.tex}

    \subsection{Filtering} \vspace{-2mm} \label{ssec:experiments-filtering}
    
    For keyword matching-based pre-filtering, we used eight words: ``\jp{声}'', ``\jp{ボイス} '', ``\jp{ヴォイス}'' (voice), ``\jp{響}'' (resonance), ``\jp{音}'' (sound), ``\jp{聴}'' (listen), ``\jp{聞}'' (hear), and ``\jp{歌}'' (song).
    The threshold for the number of keyword-matching comments per video was $10$.
    
    For machine learning-based filtering, we used pre-trained BERT~\cite{devlin2018bert} model\footnote{\url{https://huggingface.co/cl-tohoku/bert-base-japanese}}. 
    We collected $32{,}453$ labels for comments, out of which $11{,}647$ were ``speech-related.'' $80$\% and $20$\% labels were used for training and evaluation, respectively.
    We attempted to improve the performance by using a keyword subset. We examined using all combinations of the subsets.
    Finally, seven different subsets and classifiers with high precision were selected. The choice of precision for the selection is to ensure the accurate extraction of ``speech-related'' comments. The average precision of the seven classifiers was $54.3$\%. In comparison, when using only the BERT-based classifier without the keyword subsets, the precision was $38.6$\%. This confirms the effectiveness of using the keyword subsets in combination, as it improves the precision.
    After training the classifiers, we classified unlabeled comments. Videos were selected if they had $10$ or more comments identified as ``speech-related'' by any of the seven classifiers.
    Hereinafter, a subset of selected videos, including $1{,}523$, was used for further processing.

    \vspace{-2mm}
    \subsection{Quality assurance}\vspace{-2mm}
    Through VAD, we obtained $54{,}610$ speech segments.
    \Fig{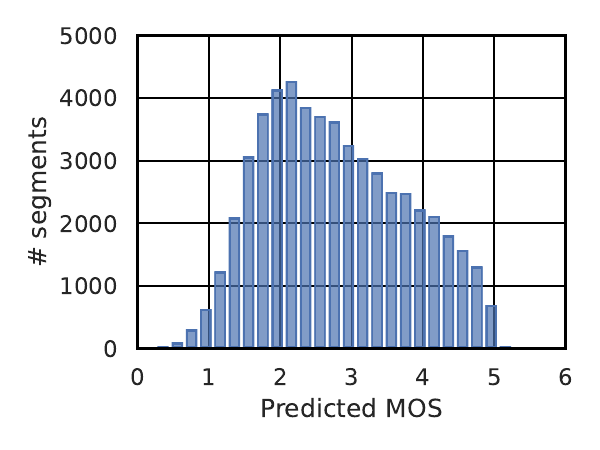} shows the distribution of NISQA-predicted mean opinion scores (MOSs) on audio quality. We set the threshold to $2$, which is the most frequent score.
    Also, segments with a duration between $2$ seconds and $10$ seconds were retained.
    The audio volume was checked using Pydub\footnote{\url{https://github.com/jiaaro/pydub}}, and segments with a volume of $-55$dB or lower were excluded.
    
    For transcription using Whisper~\cite{radford2022robust}, both the tiny and large model were employed, because the former tends to excel in fidelity to the speech while the latter excels in grammatical correctness\footnote{The average Word Error Rate (WER) of transcriptions from Whisper large model was 22.1\%. Upon final publication, we will provide manually corrected transcriptions to ensure a WER of 0.}.
    
    The NSFW detection was performed by MeCab\footnote{\url{https://taku910.github.io/mecab/}} and the Japanese NSFW dictionary\footnote{\url{https://github.com/MosasoM/inappropriate-words-ja}}.
    
    The MLM score threshold for non-verbal voice detection
    was determined to be $-0.01$. \Fig{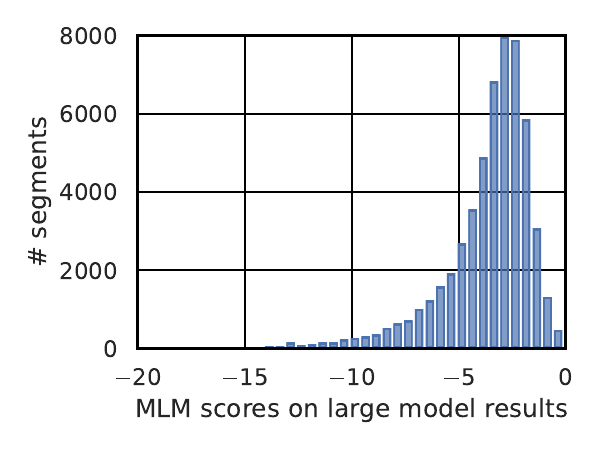} is the histogram of MLM scores calculated by the transcriptions by Whisper large model. The analysis reveals that the MLM scores of collected segments are distributed around the peak of $-3$, with a range of approximately $\pm 2$ interval. 
    We observe that the percentage of segments whose MLM scores exceed the threshold $-0.01$ was approximately $0.05$\%, which was extremely low frequency.

    \begin{figure}[t]
        \centering
        \begin{minipage}[t]{0.48\linewidth}
            \centering
            \includegraphics[width=0.98\linewidth]{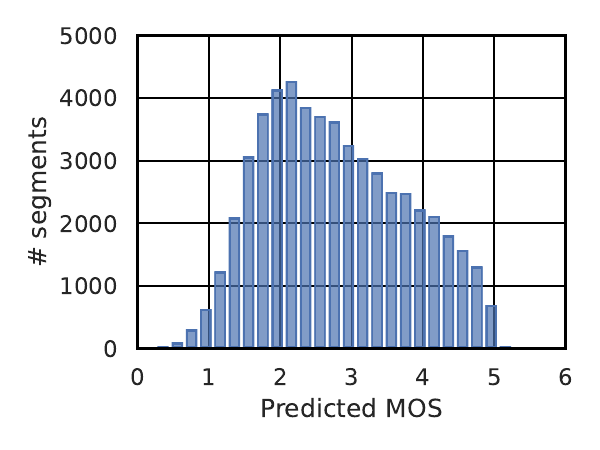}
            \vspace{-4mm}
            \caption{Histogram of NISQA-predicted MOS on speech quality.}
            \label{fig:figure/MOS_pred_hist.pdf}
        \end{minipage}
        \hfill
        \begin{minipage}[t]{0.48\linewidth}
            \centering
            \includegraphics[width=0.98\linewidth]{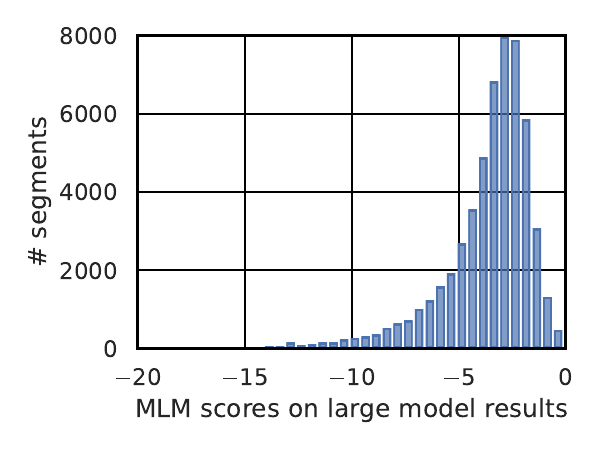}
            \vspace{-4mm}
            \caption{Histograms of MLM scores.}
            \label{fig:figure/MLM_hist.pdf}
        \end{minipage}
        \vspace{-2mm}
    \end{figure}
    
    We used the $x$-vectors extracted by xvector\_jtubespeech\footnote{\url{https://github.com/sarulab-speech/xvector_jtubespeech}} for voice characteristics variation. We performed hierarchical clustering and made $11{,}000$ clusters based on voice characteristics similarity. From each cluster, a single audio segment was randomly selected. After the selection, we further conducted manual annotation for whether the segments include NSFW words, non-target language, and multi-speakers. Finally, $7{,}667$ segments, with a total length of $30{,}661$ seconds, were selected.

    \vspace{-2mm}
    \subsection{Annotation} \vspace{-2mm}
    We hired workers through the crowdsourcing platform, Lancers\footnote{\url{https://www.lancers.jp}}. Each worker annotated 10 segments. There were a total of $1{,}318$ workers, and each worker was paid $200$ yen as reward.
    
    Before the annotation, in preparation for the machine learning experiments described at Section~\ref{ssec:experiments-baseline}, we designed the training, validation, and test sets. To avoid data leakage caused by similar voice characteristics within the same video or YouTube channel, we ensured that the sets have no overlap in YouTube channel and included a diverse range of segments. As a result, we created training, validation, and test sets with $6{,}463$, $593$ and $611$ segments respectively.
    
    We designed our corpus to include variations introduced by workers. Specifically, for the training set, we included one characteristics prompt per segment, while five prompts per segment for the other sets, following the existing studies~\cite{drossos20clotho, audiocaps}.

    \vspace{-2mm}
    \subsection{Corpus analysis}\vspace{-2mm}
    
    \begin{figure}[t]
        \centering
        \begin{minipage}[b]{0.52\linewidth}
            \centering
            \includegraphics[width=0.98\linewidth]{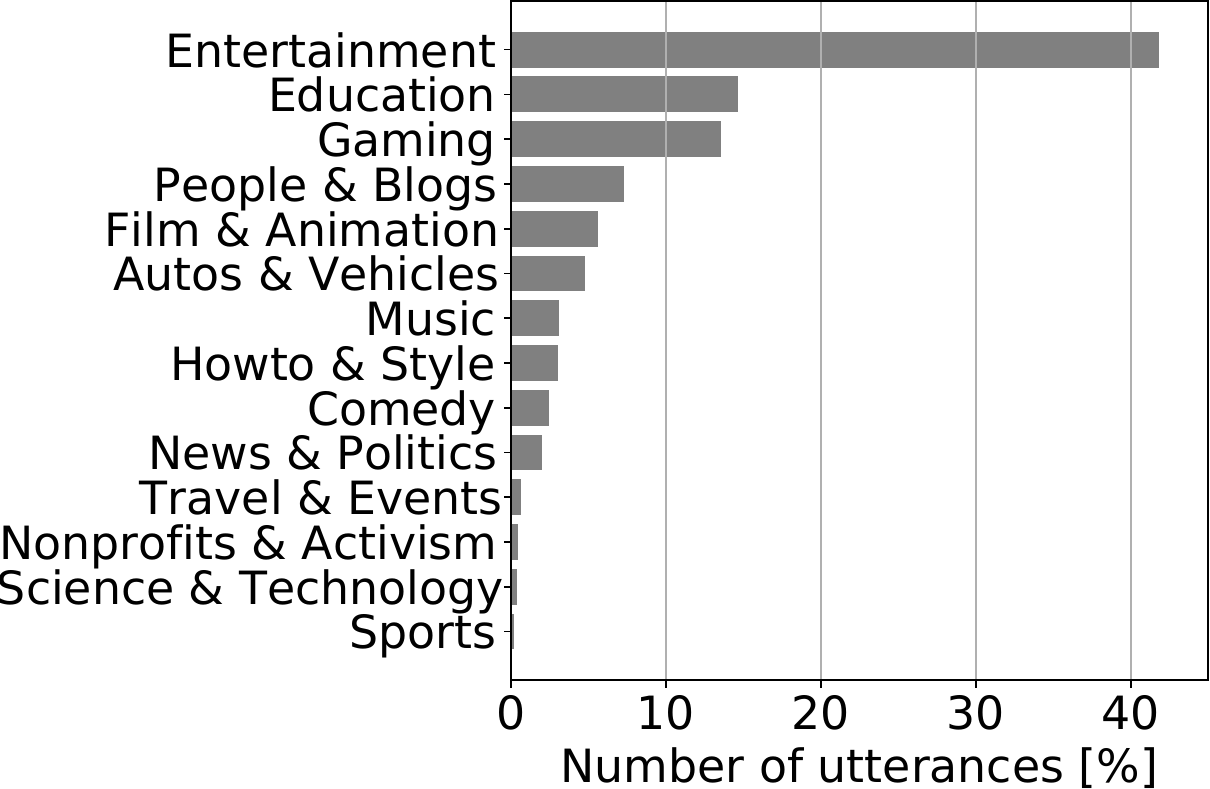}
            \vspace{-4mm}
            \caption{Categories of speech.}
            \label{fig:figure/video_category.pdf}
        \end{minipage}
        \begin{minipage}[b]{0.44\linewidth}
            \centering
            \includegraphics[width=0.98\linewidth]{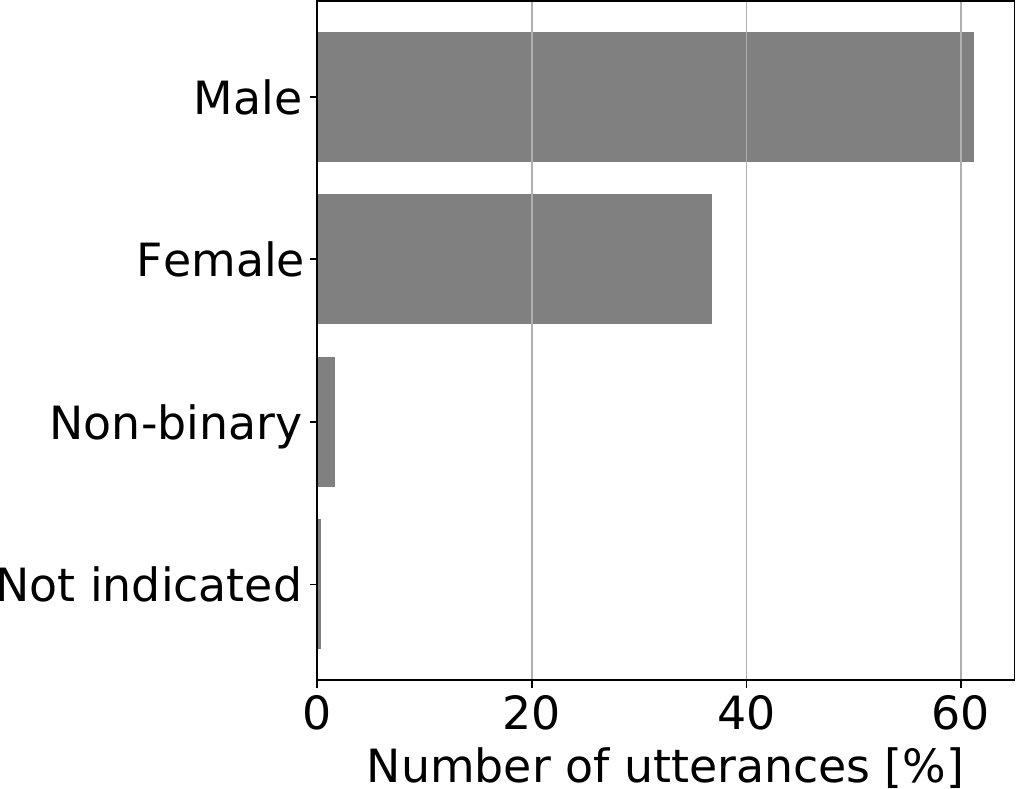}
            \vspace{-4mm}
            \caption{Gender of speech.}
            \label{fig:figure/gender.pdf}
        \end{minipage}
        \vspace{-2mm}
    \end{figure}
    
    \begin{figure}[t]
        \centering
        \begin{minipage}[b]{0.44\linewidth}
            \centering
            \includegraphics[width=0.68\linewidth]{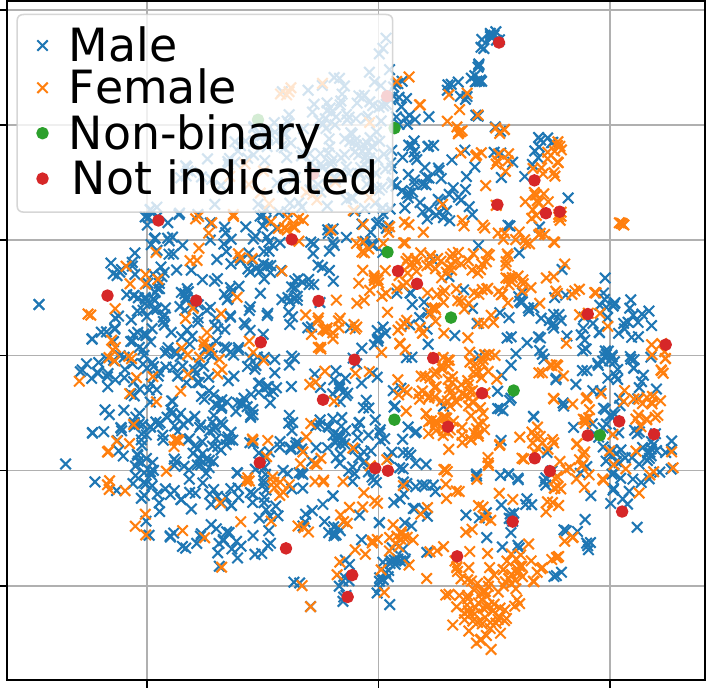}
            \vspace{-4mm}
            \caption{$x$-vector distributions colored by gender.}
            \label{fig:figure/xvector-gender.pdf}
        \end{minipage}
        \begin{minipage}[b]{0.55\linewidth}
            \centering
            \includegraphics[width=0.98\linewidth]{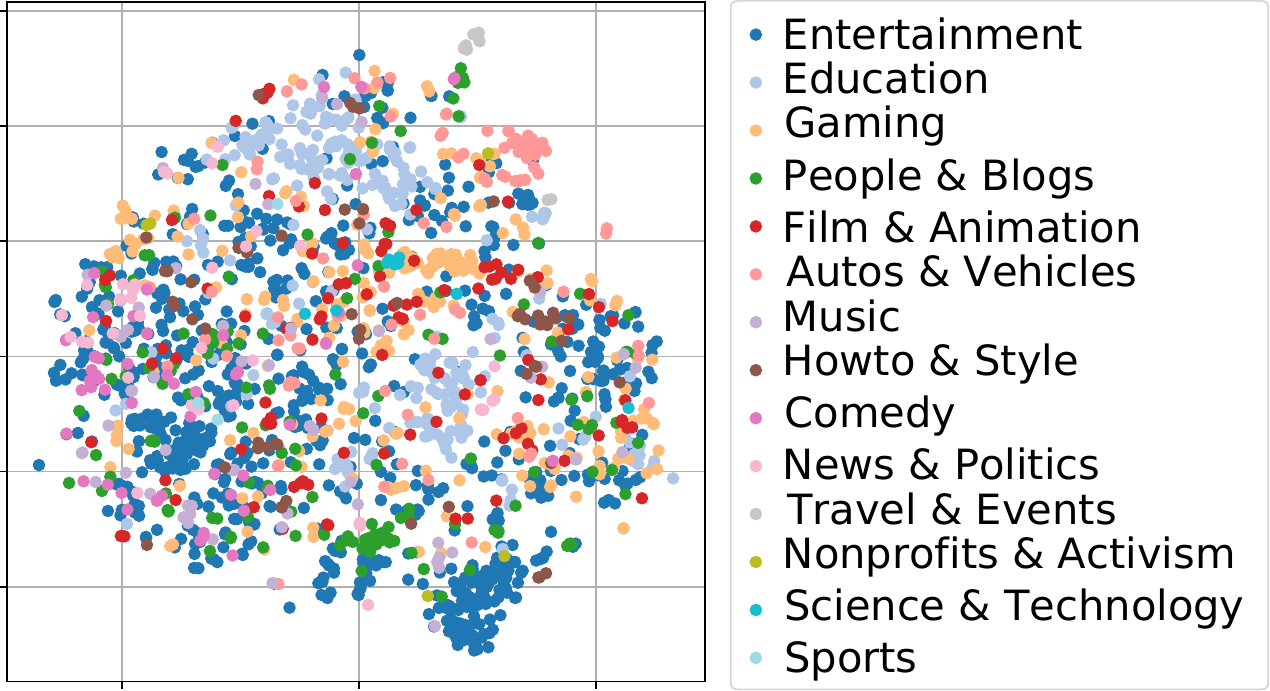}
            \vspace{-4mm}
            \caption{$x$-vector distributions colored by category.}
            \label{fig:figure/xvector-category.pdf}
        \end{minipage}
        \vspace{-2mm}
    \end{figure}
    
    
    We analyze the constructed corpus. Particularly, we investigate data diversity, which is our aim of the corpus.

    \vspace{-2mm}
    \subsubsection{Video categories} \vspace{-2mm}
    We investigated which video category speech segments in the corpus belonged to. The source video of each segment was classified according to the YouTube video category. \Fig{figure/video_category.pdf} shows the results. The corpus contains 14 categories, indicating that it covers a wide variety of categories. The top three categories (entertainment, education, gaming) account for approximately 70\%, and minor categories such as Science \& Technology are also included.
    

    \vspace{-2mm}
    \subsubsection{Gender distributions} \vspace{-2mm}
    We manually annotated gender to the characteristics prompts to analyze gender diversity. 
    \Fig{figure/gender.pdf} presents the distribution of gender. While the majority of characteristics prompts are labeled as male or female, non-binary and some prompts that don't mention gender (not-indicated) still exist. Similar to a typical TTS corpus, clusters can be observed for male and female voices. Non-binary and not indicated categories do not form distinct clusters but are scattered throughout.
    To provide a detailed analysis, we present the t-SNE visualization of $x$-vectors colored by gender in \Fig{figure/xvector-gender.pdf}. There are clusters for male and female voices. However, non-binary and not indicated categories do not form clusters but appear scattered.

    \vspace{-2mm}
    \subsubsection{Voice characteristics of video categories}\vspace{-2mm}
    To examine the relation between $x$-vector and video categories, we present the t-SNE visualization of $x$-vectors colored by the video category in \Fig{figure/xvector-category.pdf}. In the Entertainment and Education categories, specific clusters can be observed, particularly in the bottom-right and top-central regions. This suggests that typical voice characteristics are gathered within each category. On the other hand, for the majority of the scatter plot, no prominent clusters are observed. This indicates that the speech in this corpus encompass both typical voices within categories and voices that are shared across categories.

    \vspace{-2mm}
    \subsection{Machine learning baseline} \vspace{-2mm}
    Using the constructed corpus, we conduct machine learning experiments to align speech and characteristics prompts. These gives future directions of Prompt TTS.
    
    \subsubsection{Model construction} \vspace{-2mm}
    
    \drawfig{t}{0.9\linewidth}{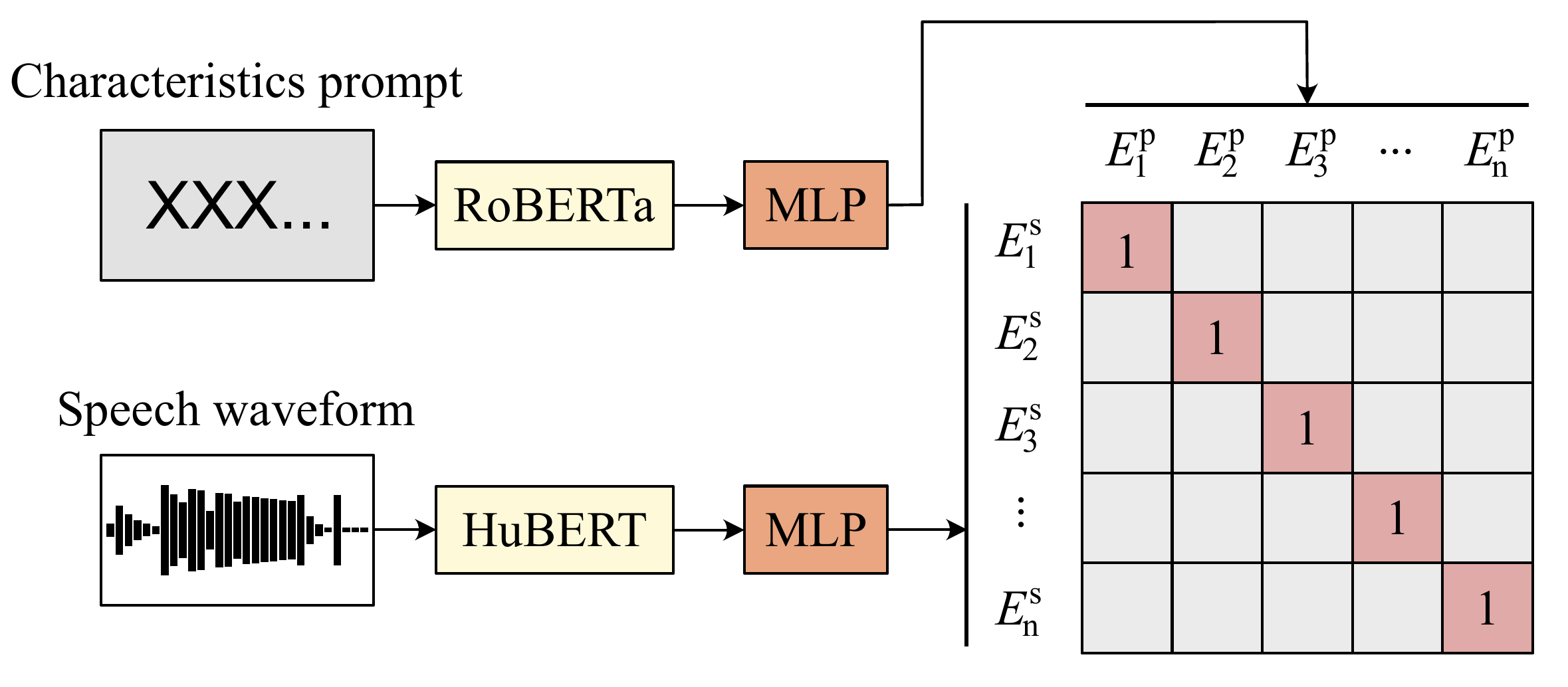}{Baseline model architecture. MLP means multi-layer perceptron. $E^s$ and $E^p$ mean the $n$-dimensional embeddings of speech and characteristics prompt, respectively.}
    \input{table/zeroshot-classification}

    \label{ssec:experiments-baseline}
    We constructed a baseline model that aligns speech and characteristics prompts.
    The model was inspired by CLAP~\cite{wu2023large}, the model that embeds both audio and text into the same space by contrastive learning. While HTS-AT~\cite{chen2022hts} is used as the audio encoder for CLAP, we changed it into HuBERT~\cite{hsu2021hubert} to grasp speech features well. We used japanese-roberta-base\footnote{\url{https://huggingface.co/rinna/japanese-roberta-base}} and japanese-hubert-base\footnote{\url{https://huggingface.co/rinna/japanese-hubert-base}} as pre-trained models of RoBERTa~\cite{liu2019roberta} and HuBERT, respectively. 
    \Fig{figure/clap.pdf} shows the overview of baseline model architecture.
    Most of hyperparameters followed official implementation of CLAP\footnote{\url{https://github.com/LAION-AI/CLAP}}. The batch size was set to $48$, and the learning rate was $0.0001$.
    We used $8$ GPUs, NVIDIA A100 for NVLink 40GiB HBM2. The training process took approximately 1 hour.

    \subsubsection{Evaluation tasks}
    Following the CLAP paper~\cite{elizalde22clap}, we evaluate the trained model and obtained embeddings.
    
    \textbf{Speech retrieval from characteristics prompt.}
    We calculate the cosine similarity between the embeddings of the input prompt and the set of embeddings of target speech segments. A higher cosine similarity value indicates a higher-ranked retrieval result. We evaluate whether the proper segment can be retrieved by the prompt.

    \textbf{Zero-shot speech classification.}
    We automatically generate characteristics prompts using categorical labels, such as ``a voice of [label].'' Then a prompt closest to the audio segment in the embedding space is selected.  The label associated with that prompt is considered as the classification label for that speech. We evaluate whether the correct label can be obtained without additional training.
    
    \subsubsection{Objective evaluation} \vspace{-2mm}
    
    We evaluated our model using mean average precision at top 10 retrieval (mAP@10) following~\cite{elizalde22clap}. mAP@10 is an evaluation metric that measures how accurately the speech corresponding to each characteristics prompt is retrieved within the top 10 retrievals. 
    At the best epoch, the text to speech mAP@10 on the test set reached $8.63$\%, while it was $0.54$\% before the training. 
    In comparison to what was trained specifically for environmental sounds~\cite{elizalde22clap}, the obtained value may appear lower. However, it is important to note that mAP@10 will be $10$\% when the 10th candidate in every retrieval is the correct pair. Therefore, an $8$\% value can be considered a reasonable indication of learning to a certain extent.
    
    To test whether the model recognizes the simple characteristics of the speech with the unseen data, we conducted the gender classification on JVS~\cite{takamichi2020jsut} parallel100 set, which consists of 49 male speakers and 51 female speakers, with each speaker having 100 speech samples. Using labels of two gender, we made two characteristics prompts: \jp{男性の声} and \jp{女性の声}, which meant ``a male voice'' or ``a female voice'' and retrieved one prompt closest to the JVS speech in the embedding space. The gender of the retrieved prompt is considered as the gender of the speech. For example, if a male speaker's speech retrieved the prompt ``a male voice,'' then the classification would be correct.
    Table~\ref{tab:zeroshot-classification} shows the confusion matrix of the result. It is observed that both genders' data are correctly identified at around $70$\%, indicating that the model had effectively learned to associate the speech of one gender with the text indicating same gender.

    \vspace{-2mm}
    \subsubsection{Subjective evaluation} \vspace{-2mm}
    \drawfig{t}{\linewidth}{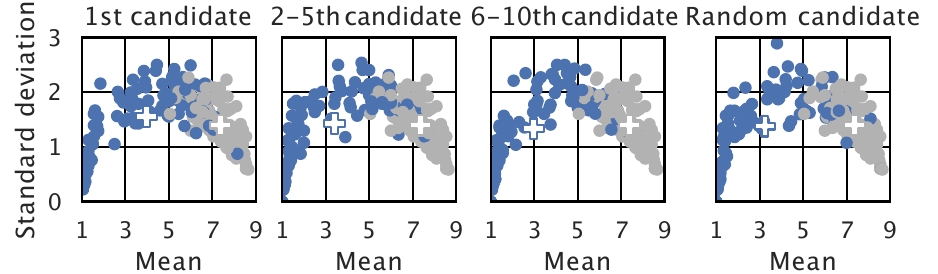}{Mean and standard deviation of subjective evaluation on each prompt-speech pair. Blue circles indicate retrieved pairs of the figure title, and gray ones indicates ground-truth. ``$+$'' marks indicate average mean and standard deviation of same color plots.}
    \begin{figure}
        \centering
        \begin{minipage}[t]{0.3\linewidth}
            \centering
            \includegraphics[width=0.98\linewidth]{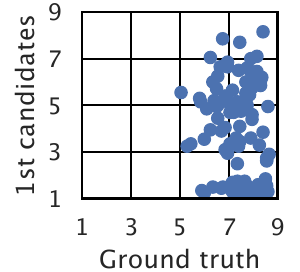}
            \caption{Mean relationship of ground truth results and 1st retrieved results, both of the same characteristics prompt.}
            \label{fig:figure/mean_GT-1.pdf}
        \end{minipage}
        \hfill
        \begin{minipage}[t]{0.66\linewidth}
            \centering
            \includegraphics[width=0.98\linewidth]{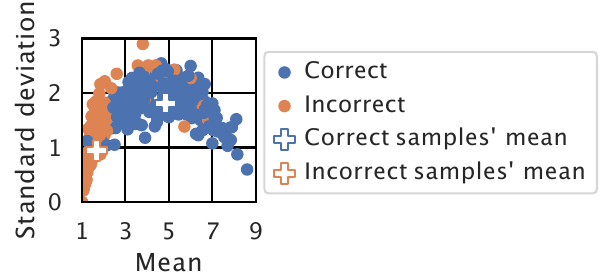}
            \caption{Mean and standard deviation of subjective evaluation on each of prompt and retrieved speech, colored by whether the retrieved speech posses gender described in the prompt.}
            \label{fig:figure/mean-stddev_gender-matching.pdf}
        \end{minipage}
    \end{figure}

\input{table/retrieved_pair_example_1column}

    To perceptually evaluate the retrieved speech by the prompt, we conducted subjective evaluations.
    
    We randomly selected $100$ characteristics prompts from the test set and retrieved speech from the whole test set. 
    From retrieval results, we created four kinds of speech paired with the characteristics prompt: 1st candidate, 2--5th candidates, 6--10th candidates, and random candidates. The first means the 1st candidate retrieved by the prompt, and the last is randomly selected from the test set.
    
    We presented the prompt and speech to crowdworkers and let them evaluate how much the prompt represents the speech characteristics on a nine-point scale, 9 is the best matching and 1 is the worst. For comparison, we added ground-truth speech (paired with the prompt in the test set) to the listening test set.
    Each worker was presented with a total of $20$ pairs. We employed $500$ workers and obtained $20$ evaluations for each pair.

    \Fig{figure/mean-stddev.pdf} shows the results. Each figure illustrates mean and standard deviation of the scores of each prompt-speech pair. Those of ground-truth pairs are illustrated for comparison. 

    \textbf{Q. Is free-form text truly appropriate in describing the voice characteristics?}
    We validate the adequacy of the free-form characteristics representation we present in this paper. As shown in \Fig{figure/mean-stddev.pdf}, the ground-truth pairs obtained a sufficiently high average score of $7.37$ despite the difference between the writers of the free-form expressions and the evaluators. It indicates that the free-form expressions can appropriately represent voice characteristics regardless of the writers or evaluators. Note that, compared to conventional categorization (e.g., gender), the scores tend to be more variable.

    \textbf{Q. Does the baseline model retrieve perceptually good speech from the given prompt?}
    We compared between retrieval rank (1st, 2--5th, 6--10th candidates and random). The average scores for each method were $3.98$ (1st), $3.42$ (2--5th), $2.98$ (6--10th) and $3.25$ (random). Statistically significant differences were observed between the 1st candidate and random ($p < 0.05$), indicating that the model can retrieve appropriate speech. However, there is still room for improvement in the trained model to reach the ground-truth score, and the samples in the 6--10th positions fall below the random method, indicating the need for improvement in the retrieval method.

    \textbf{Q. Is the low score for 1st candidate due to the low ground-truth score?}    
    As mentioned above, the scores of the ground-truth samples vary among the samples. We investigated whether this variability in ground-truth scores affects the retrieval performance of the model. \Fig{figure/mean_GT-1.pdf} illustrates the ground-truth scores along with the 1st candidate's corresponding to each prompt. From this figure, there is no clear correlation between the two, indicating that the variability in ground-truth scores has little impact on the retrieval performance. Therefore, the low scores of the 1st candidate primarily will reflect the performance of the retrieval model itself.

    \textbf{Q. What happens when scores are extremely low?}
    Retrieved candidates includes extremely low scores as observed in the bottom left of \Fig{figure/mean-stddev.pdf}. To investigate the reason behind this, we examined the correspondence between the gender of the prompt and the gender mentioned in the ground-truth prompt associated with the retrieved speech. As shown in \Fig{figure/mean-stddev_gender-matching.pdf}, in the samples with low scores, it frequently occurs that the gender is misaligned. For example, there are cases where the input prompt includes the term ``female,'' but our model retrieves a male voice. To address this issue, we need a training method that embeds the same gender samples close.

    \textbf{Q. What are the actual examples?}
    Finally, we provide examples of the input prompt and the corresponding ground-truth prompt for the retrieved speech in Table~\ref{tab:retrieved-pair-examples}. In the case of high scores ($8.15$), we can observe that not only the gender and age (``young women'') but also the style (``sweet'') are aligned. As mentioned earlier, when the gender is different, the scores significantly drop ($1.05$). On the other hand, even when the gender is aligned, if there are differences in the age group (``middle-aged'' vs. ``young'') or style (``questioning manner'' vs. ``excited''), we can see the low score ($2.35$).

%% file: table/data-collection.tex
\begin{table}[t]
\centering 
\caption{Results of data collection} \footnotesize
\begin{tabular}{l|l}
Retrieved item               & Value          \\ \hline
\#article-categories                 & $180$ categories \\
\#search-phrases             & $0.10$M phrases  \\
\#videos found in the search & $1.14$M videos  \\
Audio duration               & $0.30$M hours \\
\#comments                   & $24.2$M sentences \\
\end{tabular}
\label{tab:data-collection}
\vspace{-4mm}
\end{table}

%% file: table/zeroshot-classification.tex
\begin{table}[t]
\centering
\caption{Zero-shot gender classification}
\begin{tabular}{c c|c c}
    \multicolumn{2}{c|}{}& \multicolumn{2}{c}{Classification result} \\
    \multicolumn{2}{c|}{}& Male & Female \\
    \hline
    Actual & Male & 3442 & 1456 \\
    gender & Female & 1048 & 4051
\end{tabular}
\label{tab:zeroshot-classification}
\end{table}

%% file: table/retrieved_pair_example_1column.tex
\begin{table}[t]
\centering 
\caption{Retrieved pair examples. ``Score'' means the average of subjective evaluated appropriateness. ``Rank'' indicates the rank of retrieved candidate. Text of retrieved candidate means one characteristics prompt of retrieved candidate. Each text has been translated to English.}
\footnotesize
\begin{tabular}{
p{0.3\linewidth} p{0.06\linewidth}|p{0.06\linewidth} 
p{0.3\linewidth} p{0.06\linewidth}}
 Retrieval text & Score & Rank & Retrieved candidate & Score \\
 \hline
 A young woman in her twenties is speaking slowly with a sweet, clinging voice. & 8.40 &
 1st & A young woman is speaking in a cloyingly sweet tone. & 8.15 \\
 \hline
 A young woman in her twenties is speaking with a cheerful voice. & 8.10 & 
 1st & A young man is speaking in a gentle voice, whispering softly. & 1.05 \\
 \hline
 A middle-aged cheerful man is speaking in a clear voice, addressing in a questioning manner. & 7.75 &  
 2--5th & A young man is speaking in a high-pitched voice, as if he is excited. & 2.35 \\
\end{tabular}
\label{tab:retrieved-pair-examples}
\end{table}

%% file: section/conclusion.tex
\vspace{-3mm}
\section{Conclusion} \vspace{-2mm} \label{sec:conclusion}
In this paper, we developed a paired corpus of speech and characteristics prompts and conducted evaluations of both the corpus itself and a baseline model. 
This corpus will promote research on Prompt TTS, where the speaker is controlled by characteristics prompts. 
The consideration of Prompt TTS architecture and the expansion of the corpus itself are tasks for future work.